\documentclass[12pt,epsfig]{article}
\usepackage{amssymb}

\begin{document}
\def\be{\begin{equation}}\def\ee{\end{equation}}\def\l{\label}
\def\0{\setcounter{equation}{0}}\def\b{\beta}\def\S{\Sigma}\def\C{\cite}
\def\r{\ref}\def\ba{\begin{eqnarray}}\def\ea{\end{eqnarray}}
\def\n{\nonumber}\def\R{\rho}\def\X{\Xi}\def\x{\xi}\def\la{\lambda}
\def\d{\delta}\def\s{\sigma}\def\f{\frac}\def\D{\Delta}\def\pa{\partial}
\def\Th{\Theta}\def\o{\omega}\def\O{\Omega}\def\th{\theta}\def\ga{\gamma}
\def\Ga{\Gamma}\def\t{\times}\def\h{\hat}\def\rar{\rightarrow}
\def\vp{\varphi}\def\inf{\infty}\def\le{\left}\def\ri{\right}
\def\foot{\footnote}\def\ve{\varepsilon}\def\N{\bar{n}(s)}\def\cS{{\cal S}}
\def\k{\kappa}\def\sq{\sqrt{s}}\def\bx{{\bf x}}\def\La{\Lambda}
\def\bb{{\bf b}}\def\bq{{\bf q}}\def\cp{{\cal P}}\def\tg{\tilde{g}}
\def\cf{{\cal F}}\def\bN{{\bf N}}\def\Re{{\rm Re}}\def\Im{{\rm Im}}
\def\K{{\cal K}}\def\ep{\epsilon}\def\cd{{\cal d}}\def\co{\hat{\cal
O}} \def\j{{\h j}}\def\e{{\h e}}\def\F{{\bar{F}}}\def\cn{{\cal N}}
\def\P{\Phi}\def\p{\phi}\def\cd{\cdot}\def\L{{\cal L}}\def\U{{\cal U}}
\def\Z{{\cal Z}}\def\ep{\epsilon}\def\a{\alpha}\def\ru{{\rm u}}
\def\vep{\varepsilon}

\begin{center}
{\Large\bf  Symmetries, Variational Principles and Quantum Dynamics}

\vskip 0.5cm

{\large J.Manjavidze}\\{\it Institute of Physics, Tbilisi, Georgia
$\&$ JINR, Dubna, Russia, e-mail: joseph@nusun.jinr.ru},\\ {\large
A.Sissakian}\\{\it JINR, Dubna, Russia, e-mail: sisakian@jinr.ru}
\end{center}
\vskip 1cm

\begin{abstract}\footnotesize\it

The role of symmetries in formation of quantum dynamics is discussed.
A quantum version of the d'Alambert's principle is proposed to take
into account symmetry constrains for quantum case. It is noted that
in this approach one can find, in four space-time dimensions, the
free of divergences quantum field theory.
\end{abstract}

\vskip 1cm

{\large\bf 1.} The problem of "book-keeping" of the quantum degrees
of freedom looks till now as a primary task for any modern field
theory. The point is that the non-Abelian gauge theory of Yang and
Mills, or Einstein gravity, both obey high symmetry and the symmetry
itself signifies an existence of a connection among various degrees
of freedom, i.e. hence the symmetry reduces the number of dynamical
degrees of freedom. So, the "book-keeping" means the procedure of
counting of the $independent$ degrees of freedom.

This question has a fundamental background. It is known that the
symmetry may improve solvability of the quantum problem.  The hidden
$O(4)$ symmetry solves the Coulomb problem \C{y3} and, therefore, the
semiclassical Bohr-Sommerfeld quantization rules are exact for this
problem. Another example is the $SL(2C)$ symmetry which solves the
2-dimensional sin-Gordon field theory so that the semiclassical
approximation becomes exact \C{y4}. More examples one can find in
\C{y5}. So, the symmetry actually reduces the quantum degrees of
freedom.

We should keep in mind this marvelous phenomenon although it is hard
to hope that we always deal with such simple "completely integrable
systems" \C{y6}, in which there is enough number of symmetry
constrains  to prevent perturbations. However, we have to keep in
mind such a possibility. What if, e.g., the Einstein gravity is the
purely classical theory because of its highest space-time symmetry.
Sure, this assumption looks curiously, though, it is most likely that
Einstein was thinking it is so. But there is also another side of the
"book-keeping" problem.

To note is that general quantum field, due to the constrains, can not
be treated always as the dispersive medium, unlike the
electromagnetic fields which are described as one, two, and more
photon states. In this case the quantum electrodynamics fundamental
method of spectral analysis can not be applied. An example is the
soliton quantization problem, where ordinary methods are not
effective \C{y}.

{\it One needs other formalism, which takes into account the symmetry
constrains.}

{\large\bf 2.} The modern quantum theories practically are based on
the classical Hamiltonian variational principle. It can be built into
the quantum theories keeping in mind the Bohr's "correspondence
principle". Formally, this becomes possible thanks to an outstanding
result of Dirac that the action $S(u)$ is a generator of
displacements along the $u(t)$ trajectory. Then the transition
amplitude of initial state, $|in>$, into the final one, $|out>$, is
an exponent: $<in|out>\sim \exp\{iS(u)\}$. The wave nature of quantum
dynamics is realized then as a sum over all trajectories $u(t)$.

The simplest way of definition of this sum is based on the Ferma's
principle and uses the basis of geometrical optics. A generalization
represents an equivalent to the Hamiltonian variational principle. In
this frame the dynamics is realized on the actions extremum, i.e. it
is supposed that the $physical$ trajectory $u_c(t)$ is a solution of:
\be \f{\d S(u)}{\d u(t)}=0.\l{10}\ee This approach of "book-keeping"
of deviations from $u_c$ is known as the WKB method \C{y2}.

A weak point of this variational principle, as it is known from
classical mechanics, is that it demands a cumbersome way the
(symmetry) $constrains$ are introduced into the formalism \C{y11}.
The formal problem is to extract the dynamical, i.e. independent,
degrees of freedom. In the quantum case this leads to implementation
of the Faddeev-Popov $ansatz$ \C{y13} which separates the dynamical
degrees of freedom from symmetrical ones. Unfortunately, this method
is unacceptable for non-Abelian gauge theory (in case the
corresponding Yang-Mills fields are sufficiently strong \C{y15}, e.g.
if the soliton-like excitation is considered) and for gravitational
field \C{y14}.
\\

{\large\bf 3.} We propose to use the d'Alambert's variational
principle which easily "absorbs" the constrains, as is known from the
classics. For that reason, it is of a definite value for us.

Indeed, the d'Alambert's variational principle means that {\it the
virtual deviation $\vep(t)$ does not produce any work $F(u)\vep=
\Upsilon$ since the mechanical motion is time reversible.} So, for
mechanical systems $F(u)\vep(t)=0$ and, as soon as $\vep(t)$ is an
arbitrary function, one gets a condition that sum of $all$ the forces
should be equal zero, $F(u)=0$. This equality may be treated as the
equation of motion for $u(t)$ since $F(u)$ includes also the "force
of inertia", $\sim\ddot{u}(t)$. If there are constrains in the system
considered, then they have to be included along with other forces
into $F(u)$ \C{y6}.

It should be noted here that, as follows from d'Alambert's principle,
the solution of equation of motion $u_c$ already "absorbs" all the
constrains. Therefore, it is enough to count all possible forms of
$u_c$ which are fixed by the integration constants.

Therefore, the mechanics built on the d'Alambert's variational
principle seems to be most useful for us. The connection among
various degrees of freedom, as well as constrains, are naturally
included into the prescription that the $mechanical$ motion must be
time reversible.

The time reversibility may be achieved in quantum theories, for
instance, considering loop trajectory with the transition amplitude
$R=<in|in>$. In the meantime, to define the integral over particle
momentum unambiguously, one needs to implement the
$i\ep$-prescription of Feynman.

In particle physics the $i\ep$-prescription has been chosen so that a
wave must disappear in remote future. Note that only such definition
conserves a total probability. But it is evident that such a wave
process is not time reversible. Therefore, even the simple loop
trajectory can not be considered as an example of time reversible
wave motion: due to the $i\ep$-prescription the loop amplitude is
complex, i.e. it is not a singlet of complex conjugation operator.

This example creates an impression that quantum mechanics is time
irreversible and, therefore, the d'Alambert's variational
principle can not be claimed for it. But then one may ask where is
the quantum correspondence principle?

The answer is in fact that the amplitude $R$ is not a measurable
quantity while only the product $$ \R(in,out) =<in|out><out|in> =
<in|out> <in|out>^*= |<in|out>|^2$$ is measured experimentally, i.e.
only it has to be time reversible quantity \C{y17}. Note here that to
define the amplitude $<in|out>$ the $(+i\ep)$-prescription must be
used, as well as for $<out|in>= <in|out>^*$ the
$(-i\ep)$-prescription is applied. Therefore, just $\R(in,out)$
describes the time reversible process and this removes the
contradiction with correspondence principle. Let us consider now what
this gives us.\\

{\large\bf 4.} Following Dirac, one writes: $$ \R\sim e^{iS(u_+)}
e^{-iS(u_-)}.$$ To note is that $u_+$ and $u_-$ are two completely
independent trajectories and, by definition, the total action
$\{S(u_+)-S(u_-)\}$ is defined so that it describes closed-path
motion.

To introduce the d'Alambert's variational principle one needs to
distinguish between the $physical$ trajectory $u(t)$ and the
$virtual$ deviation, $\vep(t)$, from it. Therefore, it is naturally
that $$ u_\pm(t)=u(t)\pm \vep(t)$$ and the integrations over $u(t)$
and $\vep(t)$ are taken independently \C{y17}. Noticing that (i) the
closed-path motion is already described and (ii) the end-points,
$(in, out)$, can not varied, the integration over $\vep(t)$ must be
performed with boundary conditions for initial, $t_i$, and final
$t_f$, times: $ \vep(t_i)=\vep(t_f)=0.$ Then, in the expanding in
$\vep(t)$, $$ S(u+\vep)-S(u-\vep)=\le.\le\{S(u+\vep)-S(u-\vep)\ri\}
\ri|_{\vep=0}+2 \Re\int^{t_f}_{t_i} dt \f{\d S(u)}{\d
u(t)}~\vep(t)+U(u,\vep),$$ one may omit for the moment higher
$\vep(t)$-terms noted here by $U(u,\vep)$. The first term is
important if the classical trajectory is a periodic function \C{y17}.
The resulted integral over $\vep(t)$ gives the functional
$\d$-function: \be \int \prod_t d\vep(t)
\exp\le\{2i\Re\int^{t_f}_{t_i} dt \f{\d S(u)}{\d u(t)}~\vep(t)\ri\}
={\prod_t}'\d\le(\f{\d S(u)}{\d u(t)}\ri).\l{7}\ee This equality
means that the continuum of contributions with ${\d S(u)}/{\d
u(t)}\neq0$ are canceled in the integral independently on the shape
of the function $u(t)$. It is the result of destructive interference
among divergent $e^{+iS(u_+)}$ and convergent $e^{-iS(u_-)}$ waves.
Note that this complete cancellation is actually a result of the time
reversibility.

The dynamical equilibrium, when ${\d S(u)}/{\d u(t)}=0$, is the only
surviving and in this case the integral (\r{7}) is infinite. This is
expressed by $\d$-function in (\r{7}). Therefore, integration over
virtual deviation, $\vep(t)$, leads to the Dirac $\d$-like measure
and and the later defines the complete set of contributions in the
integrals over the trajectory $u(t)$ \C{y17}.

Returning to the d'Alambert's variational principle, one can notice
that, formally, the product $\Upsilon=\vep{\d S(u)}/{\d u}$ in the
exponent in (\r{7}) represents "virtual work". But in the considered
wave process case $\Upsilon\neq0$, i.e. the quantum virtual
deviations can produce some work. However, as follows from (\r{7}),
the "physical" trajectory is defined by the classical Lagrange
equation (\r{10}) and, therefore, is the trajectory, where dynamical
equilibrium is achieved.

Thus, we get {\it a quantum version of the d'Alambert's variational
principle}. It relies on the general principle of quantum theories
that only the time reversible amplitude module is a measurable
quantity.

Furthermore, this formalism allows including external forces \C{y18}.
Thus, if $U(u,\vep)$ takes into account the higher nonlinear in
$\vep(t)$ terms, then the $strict$ path integral representation of
$\R$ is given by: \be \R=e^{-i\mathbf{K}(j\vep)}\int {\prod_t}'
du(t)\d\le(\f{\d S(u)}{\d u(t)}+j(t)\ri)e^{iU(u,\vep)}, \l{11}\ee
where the operator, \be 2\mathbf{K}(j\vep)=\Re\int dt \f{\d}{\d
j(t)}\f{\d}{\d \vep(t)},\l{12}\ee generates a quantum perturbation
series. The interaction functional $U(u,\vep)$ can be easily
discovered from the Lagrangian. At the very end one should take in
(\r{11}) the auxiliary variables $j$ and $\vep$ to be equal to zero.

So, one can generate the quantum perturbations including the
$random$ force $j(t)$ being external to the classical system. This
is an authentic indication that our formulating of quantum wave
mechanics resembles the classical mechanics built on the
d'Alambert's variational principle base.\\

{\large\bf 5.} A generalization of the representation (\r{11}) to
the field theory case is straightforward. One should simply
replace $u(t)$ by $u(x,t)$, where $x$ is the space coordinate and
$t$ is the time \C{y19, y20}. The formalism does not depends a
type of field is considered.

It must be noticed first of all that the functional $\d$-function
in (\r{11}) establishes one to one correspondence among quantum
and classical description \C{y17}: the quantum case introduces the
random force $j(x)$ into the system influence via the Gaussian
operator $\mathbf{K}(j\vep)$. Therefore, if we know the $general$
solution $u_c\neq0$, then $all$ quantum states can be counted
varying the integration constants as the dynamical variables.

This correspondence is extended so that allows involving such a
powerful method of classical mechanics as the method of
transformation of dynamical variables. It should be underlined
that this method can not be adopted on the time irreversible
quantum amplitudes level \C{y18,y19}.

The most useful variables are the (action,angle)-type ones
\C{y20,y21}\foot{In definite sense this type of variables play the
same role as the "collective coordinates" of Bogolyubov \C{21}.}.
In this terms the world line of classical system (with enough
constrains) moves over (Arnold's) hypertorus with radii being
equal to the action variable and its location is defined by the
angle variables. In the quantum case, the action of the operator
$\exp\{-i\mathbf{K}\}$ leads to the Gaussian thrilling of the
hypertorus. This vibration may be described in terms of random
fluctuation of the action and angle variables and so that quantum
perturbation gets up not in the functional space of field
$u(x,t)$.

A transformation to the special type variables, namely to a set of
the integration constants, has been proposed in \C{y18}-\C{y21}.
To note is that if there is not enough constrains, then the system
leaves the hypertorus surface and any mapping becomes meaningless.
Such a situation is realized, for example, in  quantum
electrodynamics, which is gauge invariant but this is not enough
for the hypertorus formation. In this case ordinary method of the
Faddeev-Popov $ansatz$ is effective.

It should be stressed that discussed formulating of the theory is
necessary and sufficient as well. The necessity follows from the
fact that, in the situation of general position, the dynamics
should realized on the Arnold's hypertorus \C{y17}. The
sufficiency follows from the fact that the fluctuations are
defined on the Gaussian measure and so they cover the (action,
angle) phase space completely.

We call such perturbation theory the $topological$ theory. Indeed,
the topological properties are crucial since if a trajectory on
the hypertorus has definite winding number (topological charge),
then the probability that after random quantum walking, the system
returns to the same (or shifted on a constant winding number)
position is equal to zero. Therefore, such systems have to be
exactly semiclassical. The quantitative prove of this conclusion
is given in \C{y18,y19}.

In the case if there is not the topological charge, the system
"freely" moves along the hypertorus and can be quantum. It is
indeed evident that the Yang-Mills theory in the real-time metric
\C{y21}, and could be, may be the gravitation theory, are just
such kind of theories.

The above described transformation to the (action, angle)-type
variables reduces the field-theoretical problem to the quantum
mechanical one. Generally, such a mapping is singular but it is
possible to isolate these infinities (they look like divergent
integrals) and remove them via the normalization. This observation
is extremely important since as a result we get the quantum
mechanics being theory which is free from divergences. One can
mention that isolating the divergences and then canceling them the
renormalization was already performed. This property is intriguing
since finiteness of the theory was attained without introduction
of new fields, e.g. supersymmetry partners.

It is understandable that there is not the necessity to
distinguish between the symmetry and dynamical degrees of freedom
in new formulation of theory. Indeed, the quantum fluctuations are
realized in the (action, angle) factor space and this does not
touch the action symmetry. As a result, the perturbation series
consists of transparently invariant terms \C{y21}.
\\

{\large\bf 6.} We would like to note the following fact being
important from the cognition point of view. It is remarkable that
restricting the problem and computing the module of amplitude,
i.e. leaving the phase of amplitude undefined, the remarkable
progress is shown to be achieved. So, we can say that, for
instance, the Yang-Mills theory $exists$\foot{See {\bf
www.claymath.org/prizeproblems} concerning more details for the
question of $existence$ of the Yang-Mills theory.} as the theory
of observables since in this restricted frame (i) it is free from
the divergences and (ii) there is no problems \C{y21} with
necessity to distinguish between the dynamical and the symmetry
degrees of freedom.

However, finally, it must be admitted that one does not know,
wether the time reversibility principle, i.e. the quantum
d'Alambert's principle, can be applied to the Einstein gravity? If
yes, then no gravitational waves are expected to be observed
despite the fact that the gravity would be the quantum theory and,
moreover, such theory could be free from divergences. The point is
that, most likely, the gravity does not represent the dispersive
media. This follows from the fact that the equation for metrics
has a nontrivial solution.


\end{document}